# The mitotic spindle in the one-cell *C. elegans* embryo is positioned with high precision and stability

Jacques Pecreaux[*,#,1,2], Stefanie Redemann[*,3], Zahraa Alayan[1,2], Benjamin Mercat[1,2], Sylvain Pastezeur[1,2], Carlos Garzon-Coral[4], Anthony A. Hyman[4], Jonathon Howard[5,#]

[1] IGDR, Unité Mixte de Recherche 6290, Centre National de la Recherche Scientifique, 2 avenue du Professeur. Léon Bernard, CS 34317, 35043 Rennes cedex, France

[2] IGDR, Institute of Genetics and Development of Rennes, University Rennes 1, F-35043 Rennes, France

[3] Present address: Medizinisch Theoretisches Zentrum (MTZ), Technical University of Dresden, Fiedlerstrasse 42, D-01307 Dresden, Germany.

[4] Shriram Center of Bioengineering and Chemical engineering, Stanford University, 433 Via Ortega, Stanford, CA 94305

[5] Max Planck Institute of Molecular Cell Biology and Genetics
PfotenhauerStr. 108, 01307 Dresden, Germany

[6] Department of Molecular Biophysics & Biochemistry, Yale University, 266 Whitney Avenue, New Haven, CT 06511

[*] These authors contributed equally.
[#] Corresponding author: jacques.pecreaux@univ-rennes1.fr, jonathon.howard@yale.edu

Running title: The high stability of mitotic spindle centering

# Abstract

Precise positioning of the mitotic spindle is important for specifying the plane of cell division, which in turn determines how the cytoplasmic contents of the mother cell are partitioned into the daughter cells, and how the daughters are positioned within the tissue. During metaphase in the early *C. elegans* embryo, the spindle is aligned and centered on the anterior-posterior axis by a microtubule-dependent machinery that exerts restoring forces when the spindle is displaced from the center. To investigate the accuracy and stability of centering, we tracked the position and orientation of the mitotic spindle during the first cell division with high temporal and spatial resolution. We found that the precision is remarkably high: the cell-to-cell variation in the transverse position of the center of the spindle during metaphase, as measured by the standard deviation, was only 1.5% of the length of the short axis of the cell. Spindle position is also very stable: the standard deviation of the fluctuations in transverse spindle position during metaphase was only 0.5% of the short axis of the cell. Assuming that stability is limited by fluctuations in the number of independent motor elements such as microtubules or dyneins underlying the centering machinery, we infer that the number is on the order of one thousand, consistent with the several thousand of astral microtubules in these cells. Astral microtubules grow out from the two spindle poles, make contact with the cell cortex, and then shrink back shortly thereafter. The high stability of centering can be accounted for quantitatively if, while making contact with the cortex, the astral microtubules buckle as they exert compressive, pushing forces. We thus propose that the large number of microtubules in the asters provides a highly precise mechanism for positioning the spindle during metaphase while assembly is completed prior to the onset of anaphase.

# Introduction

During cell division, the correct positioning and orientation of the mitotic spindle are important for the developmental fate of the daughter cells. This is because the cleavage furrow usually bisects the spindle (1, 2) and thereby determines, in part, how the cytoplasmic contents are distributed to the two daughter cells (3-5). The plane of cell division also specifies the location of the daughter cells within the tissue (6). The initial establishment of spindle position and orientation in the early phases of mitosis are thought to be due to the microtubule-dependent motor protein dynein acting at the cell cortex (7) and/or in the cytoplasm (8-10). Once the spindle reaches the cell center, its position and orientation must be precisely maintained during metaphase (11, 12) until the spindle assembly checkpoint is passed and the cell enters anaphase, when chromosome segregation occurs.

In this work, we have asked: following the initial positioning of the spindle at the cell center early in mitosis, how accurately, precisely and stably is the position maintained during metaphase? By accuracy, we mean how close, on average, is the midpoint of the spindle to the center of the cell and how close on average is the orientation parallel to the anterior-posterior (A-P) axis. By precision, we mean how much variability is there from cell to cell? And by stability, we mean how well do individual cells maintain their spindle position and orientation during metaphase.

These are important questions because the reliability of biological processes ultimately depends on the number of molecules involved. The statistical fluctuations in the number of molecules often follows a Poisson distribution in which the variance is proportional to the mean (see e.g. (13)). If this holds true for the centering machinery, the standard deviation of the number will be proportional to the mean number of motors and the relative fluctuations will be inversely proportional to the square root of the number of constituent motors. This result holds independent of viscous properties of the cytoplasm, which will determine the timescale, but not the amplitude, of the fluctuations. If the motors are not independent of each other (i.e. they tend to operate in groups due to elastic or viscous coupling) or there are other sources of "noise" (such as Brownian motion) then the relative fluctuations will be larger. Thus, the number of constituent molecules places an upper limit on the precision and stability of a process. Physical and genetic perturbations indicate that the spindle is maintained at the cell center by a force-generating machinery that relies on the astral microtubules, which grow out from the spindle poles towards the cell cortex (14, 15). Thus, measurements of the accuracy and stability of spindle position may allow us to estimate the minimum number of microtubules and/or motors (e.g. dynein) that are involved in maintaining the spindle at the cell center.

We have used the one-cell embryo of the nematode *C. elegans* as a model system to study the precision of centering because the anatomy of the spindle is well characterized and its large size facilitates the tracking of spindle position. Furthermore, there is a clearly defined period of about 2 minutes, roughly corresponding to metaphase, when the spindle is relatively quiescent and statistical measurements can be made.

## Material and Methods

**Culturing C. *elegans***

C. *elegans* embryos were cultured as described in (16). The TH65 (YFP::TBA-2, also denoted YFP::α-tubulin) and TH66 (EBP2-2::GFP) worm strains used for the microtubule landing assay were maintained at 25 °C, TH27 (TBG-1::GFP, also denoted γTUB::GFP), TH30 (γTUB::GFP, histone H2B::GFP), TH290 (*gpr-1(ok2126)* back-crossed 9 times), TH291 (*gpr-2(ok1179)* back-crossed 10 times) were maintained at 16 °C. The transgenes encoding the GFP and YFP fusion proteins are under the control of the *pie-1* promotor. Transgenic worms were created by microparticle bombardment (BioRad), as described in (17).

**Gene silencing by RNA interference**

RNAi experiments were performed by feeding or injection as described in (18). The feeding clones for *zyg-9* were ordered from Gene Service: the target gene was subcloned into the RNAi feeding vector L4440 and transformed into HT115 (DE3) bacteria. Worms were grown for 4 hours at 25 °C on the plates. For injections, a region from the gene was amplified by PCR using N2 genomic DNA as a template. The PCR-sample was subsequently purified using the Qiagen PCR cleanup Kit. For T3 and T7 transcription, the Ambion kit was used and was purified using the RNeasy kit. Primers used to amplify regions from N2 genomic DNA for dsRNA production were

*gpr-1/2*: T7:   TAATACGACTCACTATAGGTCAGCGGTTGTTTTATTGAAGAT
          T3:   AATTAACCCTCACTAAAGGTGGACGAGCTGGAAAAATATAAA
*lin-5:*   T7:   TAATACGACTCACTATAGGCGAGCAAAGAAGTCTGGAGG
          T3:   AATTAACCCTCACTAAAGGCGTTCCTCTCTTCGTCAAGG
*nmy-2*   T7:   TAATACGACTCACTATAGGAATTGAATCTCGGTTGAAGGAA
          T3:   AATTAACCCTCACTAAAGGACTGCATTTCACGCATCTTATG
*lov-1*   T7:   TAATACGACTCACTATAGGAACTCATAGGTGCCAATGCC
          T3:   AATTAACCCTCACTAAAGGGCGATTTGCTCCTACCTTGA

The time post-injection when worms were assayed were 21-26 hrs (*gpr-1/2*), 40-42 hrs (*lin-5*), 18-24 hrs (*nmy-2*) and 41 hrs (*lov-1*). The knockdown of *gpr-1/2* and *nmy-2* were partial: *nmy-2* (*RNAi*) abolished oscillations but only delayed posterior displacement and *nmy-2* (*RNAi*) preserved embryo polarity.

**Centrosome imaging, tracking and analysis**

Embryos of the control TH27 strain, *gpr-1/2(RNAi)*, *zyg-9(RNAi)* and the fixed embryos were imaged using an AxioVision imager 2e upright microscope (Zeiss, Jena, Germany). All other embryos were imaged with an AxioImager M1. The microscopes were modified for long-term imaging by adding an extra heat filter in the mercury lamp light path, and by using a 12 nm bandpass excitation filter centered on 485 nm (AHF Analysentechnik, Tübingen, Germany). These filters helped to prevent phototoxicity and reduce bleaching. Images were collected using a 512x512 pixel back-illuminated emCCD camera (iXon+ on the AxioVision and an iXon 3 on the AxioImager) from Andor Technologies (Belfast, UK) running Solis software. We confirmed that photodamage was not serious by checking that the rate of subsequent divisions was normal (19). The acquisition frame rate was 31.23 frames/s for the iXon+ 32.95 frames/s for the iXon3, and with 4096 frames (corresponding to a total time of 131 s and 124.3 s respectively), the frequency ranges were 7.6 mHz to 15.6 Hz and 8.0 mHz to 16.5 Hz s respectively. The time interval used for measurements started 30 seconds after nuclear envelope breakdown. All analysis software was written in Matlab (The Mathworks). Statistical significance calculated with a two-tailed Welch t-test.

**Data representation**

The anterior and posterior centrosomes coordinates, $(A_x, A_y)$ and $(P_x, P_y)$ were computed as described in the text. We calculated the spindle coordinates $(S_x, S_y, S_l, S_a)$ as:

$$\begin{cases} S_x = \dfrac{P_x + A_x}{2} \\ S_y = \dfrac{P_y + A_y}{2} \\ S_l = \sqrt{(P_x - A_x)^2 + (P_y - A_y)^2} \\ S_a = \mathrm{atan2}(P_y, P_x) \end{cases} \qquad 1$$

where atan2 is the four-quadrant inverse tangent (MatLab).

**Cell-cycle timing**

We used the fluorescent images to define the stages of the cell cycle. Taking advantage of the dim cytoplasmic labeling by the γTUB::GFP, nuclear envelope breakdown (NEBD) was defined as the minimum of the fluorescence intensity of the nucleus (measured at the midpoint between centrosomes) (20). Anaphase onset was defined as the midpoint of the inflection in spindle elongation. The latter criterion was checked using a γTUB::GFP histone H2B::GFP line, in which we compared our estimate with the chromatid separation timing. The difference between the spindle elongation inflection and the onset of chromatid separation was $10 \pm 5$ s ($N = 4$, $p = 0.16$ compared to no difference).

**Power spectra and curve fitting**

From the time-series of the spindle's transverse position, we computed the one-sided power spectral density function, $\tilde{G}(f)$, where $f$ is frequency, using MatLab's fast Fourier transform (21). We refer to this as the power spectrum. It has the property that

$$\int_0^{f_{max}} \tilde{G}(f)\,df = \Delta f \sum_{i=0}^{2048} \tilde{G}_i = \sigma^2 \qquad\qquad 2$$

where $f_{max} \approx 16$ Hz is the maximum frequency, $\tilde{G}_i$ is the value in the $i$th frequency increment, $\Delta f \approx 8$ mHz is the frequency increment and $\sigma^2$ is the variance of the time series.

We fit the data to a Lorentzian model and to second-order model. The Lorentzian model is defined by

$$H(f) = \frac{4\sigma^2 \tau}{1+(2\pi f \tau)^2} = \frac{D/\pi^2}{f_c^2 + f^2} \qquad\qquad 3$$

where $\tau$ is the time constant (or correlation time) and $\sigma^2$ is the total variance. Alternative parameters are a diffusion coefficient $D = \sigma^2/\tau$ and a characteristic frequency $f_c = 1/2\pi\tau$. The second order model is defined by

$$H(f) = \frac{4\sigma^2 \tau_2}{1+(1-2\tau_1/\tau_2)(2\pi f \tau_2)^2 + \tau_1^2/\tau_2^2 (2\pi f \tau_2)^4} \qquad\qquad 4$$

where $\tau_1$ and $\tau_2$ are the shorter and longer time constants. This reduces to the Lorentzian when $\tau_1 \ll \tau_2$.

To estimate the parameter values we performed least-square fitting, minimizing:

$$\sum_{i=1}^{2048} \frac{\left(\langle \tilde{G}_i \rangle - \tilde{H}_i\right)^2}{\tilde{H}_i^2} \qquad\qquad 5$$

where $\langle \tilde{G}_i \rangle$ is averaged power spectrum over the 8 embryos, $\tilde{H}_i$ is the theoretical spectral density (the Lorentzian or the second-order model) and $i$ indexes the different frequencies. We corrected the fit parameters for a small systematic bias (eq. 41 of (22)).

**Stability of centration by microtubule pushing**

Pushing forces generated by microtubules that grow out from the centrosome (the astral microtubules) and make contact with the cortex will lead to centration of the spindle. There are two lines of evidence that microtubules continue to grow after contact with the cortex (and therefore generate pushing forces): (i) in vivo, EB1, a marker for microtubule growth, continues to bind to microtubule ends after they contact the cortex (23), and (ii) in vitro, microtubules fixed at one end buckle when the other end makes contact with a solid surface (24), indicating that growth continues and that compressive (pushing) forces are generated.

How pushing forces lead to centering has been model by (25). In the one dimensional model of a spindle, microtubules grow to the left and to the right with speed $v_+$. After contacting the cortex, they continue to grow for a short time, $\tau_p$, during which they generate pushing forces before shrinking with speed $v_-$. If the spindle moves a distance $z$ away from the center to the right, then there will be a difference in the probability of microtubules pushing from the right compared to the left, $\Delta p(z) = p_r(z) - p_l(z)$, because the microtubules on the right side will spend less time growing out to the cortex ($\tau^+ = (R-z)/v_+$) and shrinking back ($\tau^- = (R-z)/v_-$) and a larger fraction of the time pushing at the cortex than those on the left. The probability is: $p(z) = \tau_p / (\tau_p + \tau^+(z) + \tau^-(z))$, where we have dropped the left/right subscript. This leads to a centering force:

$$f(z) = \frac{M}{2} \Delta p(z) \overline{f} \qquad 6$$

where $M$ is the total number of microtubules and $\overline{f}$ is the force exerted while pushing. Differentiating with respect to $z$, we obtain

$$\frac{d\Delta p}{dz}(0) = 2 \frac{p_0(1-p_0)}{R} \qquad 7$$

where the subscript 0 denotes the probability at the center. The centering stiffness is

$$K = \frac{df}{dz}(0) = M p_0 (1-p_0) \frac{\overline{f}}{R} \qquad 8$$

Because motor pushing is a binomial process, the force variance is

$$\sigma_f^2 = M p_0 (1-p_0) \overline{f}^2 \qquad 9$$

Therefore, the positional variance is

$$\sigma_z^2 = \frac{\sigma_f^2}{K^2} = \frac{R^2}{M p_0 (1-p_0)} \qquad 10$$

For our case where $p_0 \ll 1$, the SD of the positional fluctuation divided by the cell radius ($R$) is inversely proportional to the square root of the number of pushing microtubules:

$$\frac{\sigma_z}{R} = \frac{1}{\sqrt{M p_0}} \qquad 11$$

Importantly, (i) the positional fluctuation depends on the square root of the number of pushing microtubules, $M p_0$, as claimed in the Introduction. And (ii) the positional fluctuations are independent of the pushing force, which can be understood through a dimensionality argument:

there is only one relevant force because the thermal fluctuations are expected to be very small (15) and therefore ignored.

## Results

To measure the accuracy and stability of spindle position and orientation, we first defined a coordinate system for the one-cell *C. elegans* embryo. We labeled the centrosomes with γTUB::GFP (Figure 1A, see Methods) and used a tracking algorithm based on (26) to locate the centroids of the anterior and posterior centrosomes (Figure 1A, red and blue traces respectively). To relate the positions of the centrosomes to the geometry of the cell, we used an active contour algorithm (27) to locate the cell periphery; the algorithm was applied to an optical plane that included the two centrosomes, and the cytoplasmic γ-tubulin fluorescence marked the interior of the cell. The cell center was defined as the centroid of the perimeter. We used the zeroth, first and second moments of cell area (i.e. within the perimeter) to determine an elliptical model of cell (28). The long axis of the ellipse (drawn through the cell center) defined the anterior-posterior axis, abbreviated AP, and was used as the *x*-axis of our coordinate system, with positive towards the posterior (Figure 1A, magenta). The short axis of the ellipse (also drawn through the cell center) defined the transverse axis and was used as the *y*-axis (Figure 1A, green). The lengths of the long and short axes of the cells were $51.3 \pm 1.7$ μm and $32.5 \pm 1.1$ μm ($N = 28$ embryos; errors are SD unless stated otherwise). Using this coordinate system, we measured the center of the spindle (defined by the mid-point of the line connecting the centrosomes), the orientation of the spindle and the length of the spindle (Figure 1B).

**Accuracy and precision of centering**

To determine the accuracy of centering, we tracked the positions of the centrosomes and computed the coordinates of the spindle for 8 minutes from before nuclear envelope breakdown (NEBD, time zero indicated by the left-hand vertical dashed line in Figure 1C-F) to the end of anaphase. After fertilization and the meeting of the male and female pronuclei in the posterior half of the embryo, the pronuclei-centrosome complex moves to the cell center. At NEBD, the anterior and posterior centrosomes lay close to the AP-axis (Figure 1C,D), the spindle center lay close to the cell center (Figure 1E) and the spindle was oriented approximately parallel to the AP-axis (Figure 1F, $S_a$). About 100 s after NEBD, the posterior centrosome began to move towards the posterior (Figure 1D, $P_x$), leading to a posterior displacement of the spindle center (Figure 1E, $S_x$). The spindle continued to lie on the AP-axis until anaphase onset when the spindle started rocking, which is best seen in the transverse position of the posterior centrosome (Figure 1D, $S_y$) and the spindle orientation (Figure 1F, $S_a$).

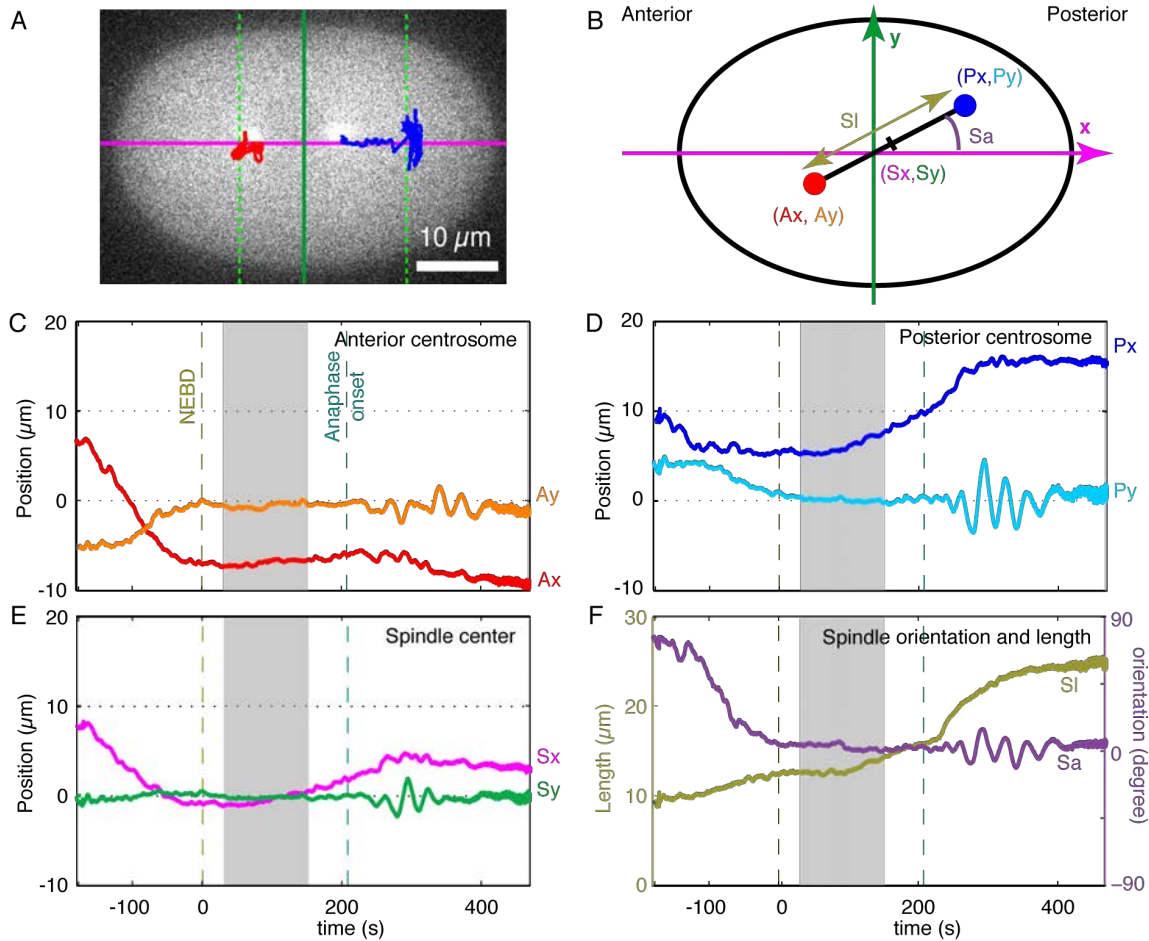

**Figure 1: Definition of spindle coordinates and typical trajectories in control embryos**
**A** Trajectories of the anterior (red) and posterior (blue) poles superimposed on a fluorescence micrograph of a one-cell C. *elegans* embryo labeled with GFP-gamma-tubulin. The horizontal magenta line is the AP-axis, and the vertical green line is the transverse axis. The intersection of the axes defines the cell center. Vertical green dashed lines mark the position of centrosomes on the AP-axis when anaphase ends. **B** Definition of the spindle coordinates and the color scheme used in the subsequent panels. **C** *x*- (orange) and *y*-coordinates (red) of the anterior centrosome of a typical cell. The dashed maroon line marks nuclear envelope break down (NEBD) and the dashed green line marks anaphase onset. The gray zone marks the maintenance phase, during which the transverse position and the orientation are stationary, and which is analyzed in detail. **D** *x*- (light blue) and *y*-coordinates (dark blue) of the posterior centrosome. **E** *x*- (magenta) and *y*-coordinates (green) of the spindle center. **F** Spindle length (khaki) and orientation (violet).

For detailed analysis, we chose the approximate 2-minute interval from 30 s to 160 s after NEBD, indicated by the gray shading in Figure 1C-F. During this interval, termed the maintenance phase (29, 30), the metaphase plate is established and maintained, the centrosomes remained stably centered on the AP axis, and there is little drift in the transverse direction. For 30 embryos, the mean and SD of the coordinates of the spindle center were -1.32 ± 0.99 μm along the AP axis (SE = 0.18 μm). These data are displayed in Table 1. The mean displacement from the center was 2.6% of the long axis of the cell in the anterior half of the embryo; this displacement, though small, as noted by others (12, 31), is significantly different from 0 (*t*-test, *p* = 5×10$^{-8}$). The mean and standard deviation of the coordinates of the spindle center were -0.05 ± 0.47 µm (*N* = 30) along the transverse axis. A zero mean position along the transverse axis was expected because embryos were imaged in an arbitrarily oriented plane that included the AP-axis. The SD, however, contains information: the value is 1.5% of the short axis of the cell and indicates that centering has high precision (i.e. the variation from cell to cell is small). The mean and SD of the spindle angle relative to the AP axis was 1.8 ± 6.3 degrees (*N* = 30); the small SD again indicates high precision. Thus, the accuracy and precision of spindle centering in the one-cell *C. elegans* embryo are high.

**Table 1: The accuracy and precision of centration.** The mean for $S_x$ gives the axial accuracy and the SDs give the axial ($S_x$), transverse ($S_y$) and angular ($S_a$) precision. Note that transverse and angular accuracy could not be determined due to symmetry (there is no maker to distinguish dorsal from ventral).

| Condition | $S_y$ (µm) | $S_x$ (µm) | $S_a$ (degrees) |
|---|---|---|---|
| Control (*N* = 30) | -0.05 ± 0.47 | -1.32 ± 0.99 | 1.85 ± 6.29 |
| Low-drift control (*N* = 8) | 0.02 ± 0.27 | -0.88 ± 1.08 | 1.58 ± 2.75 |
| *gpr-1/2(RNAi)* (*N* = 8) | 0.11 ± 0.58 | 1.08 ± 1.98 | -7.05 ± 15.75 |
| *zyg-9(RNAi)* (*N* = 8) | 0.01 ± 0.36 | -0.25 ± 1.27 | -17.23 ± 29.84 |

**The stability of spindle positioning**

To assess the stability of the spindle centering machinery, we measured the variance of the fluctuations of spindle position and orientation in individual embryos. We band-pass filtered the time-traces between 0.1 Hz and 1.1 Hz using the robust local regression algorithm (rloess) (21) to remove mean, residual drift (see below), avoid any contribution from spindle oscillations (0.04 Hz, (26)) and remove high-frequency noise due to the tracking algorithm (see below). The standard deviations of these filtered traces were 35.8 ± 18.0 nm (mean in quadrature ± SD, *N* = 30) along the AP-axis ($S_x$) and 22.8 ± 3.6 nm along the transverse axis ($S_y$) (Table 1, top line). These data are displayed in Table 2. The standard deviation of the filtered orientation ($S_a$) was 0.26 ± 0.07 degrees. For comparison, the filtered standard deviation of the transverse fluctuations for a methanol-fixed cell was 8.6 nm; the corresponding AP standard deviation was 9.6 nm. Thus, the filtered SDs, though larger than the measurement noise, are nevertheless small and indicate that the stability of positioning is very high.

**Table 2: Stability of centering in the frequency range 0.1 to 1 Hz.** The standard deviation (mean in quadrature ± SD) of position along the AP axis ($SD_x$) and along the transverse axis ($SD_y$) and of angle ($SD_a$) in the frequency range 0.1 to 1 Hz. The homologous vertebrate protein is given in column 2.

| Condition | Vertebrate protein | $SD_y$ (nm) | $SD_x$ (nm) | $SD_a$ (degrees) |
|---|---|---|---|---|
| Control ($N$ = 30) | - | 22.8 ± 3.6 | 35.8 ± 18.0 | 0.26 ± 0.07 |
| Low-drift control ($N$=8) | - | 22.1 ± 2.9 | 30.4 ± 3.8 | 0.24 ± 0.04 |
| *gpr-1/2(RNAi)* ($N$ = 8) | LGN | 18.4 ± 2.6 | 19.4 ± 3.5 | 0.20 ± 0.02 |
| *zyg-9(RNAi)* ($N$ = 8) | chTOG/XMAP215 | 39.8 ± 17.9 | 41.3 ± 10.4 | 0.45 ± 0.17 |
| *gpr-1(ok2126)* ($N$ = 5) | LGN | 30.8 ± 4.0 | 33.8 ± 3.0 | 0.37 ± 0.05 |
| *gpr-2(ok1179)* ($N$ = 5) | LGN | 25.3 ± 4.8 | 29.6 ± 2.8 | 0.27 ± 0.03 |
| *lin-5(RNAi)* ($N$ = 6) | NUMA | 23.2 ± 8.3 | 23.8 ± 6.3 | 0.24 ± 0.06 |
| *lov-1(RNAi)* ($N$ = 6) | PKD1 | 25.2 ± 1.5 | 31.5 ± 3.0 | 0.25 ± 0.03 |
| *nmy-2(RNAi)* ($N$ = 11) | non-muscle myosin | 45.8 ± 16.3 | 42.5 ± 8.8 | 0.64 ± 0.35 |

**Spindle fluctuations**

To evaluate the stability of centering over a broad frequency range, we performed Fourier analysis of the transverse position of the spindle center during the maintenance phase (e.g. Figure 1E, $S_y$). We focussed our attention on the transverse position because the axial position of the spindle is not stationary: it moves along the AP axis towards the posterior pole during metaphase (Figure 1E, $S_x$). We selected for detailed study a subset of eight of the thirty embryos that had low drift along the transverse axis. The reason for this selection was that we attribute the drift, which was typically less than 0.5 μm but none-the-less large compared to the standard deviation of the higher-frequency fluctuations, to an imbalance in the mean number of active force-generating elements above and below the AP axis; this imbalance leads to a steady displacement along the transverse axis. On the other hand, we attribute the fluctuations in transverse position to be due to the temporal fluctuations in the number of active force generating elements. It is this number that we are interested in estimating from the amplitude of the fluctuations. Thus, our stability analysis is focussed on the fluctuations, rather than the mean or the slow drift. These eight embryos had similar filtered SDs to the full set of thirty embryos (30.4 ±3.8 nm, 22.1 ± 2.9 nm and 0.24 ± 0.04 for *x*, *y* and angle; mean ± SD).

For each embryo, we computed the power spectrum as the one-sided power spectral density function of the time series of the spindle position along the transverse axis, $S_y$. The power spectrum is a measure of the variance of the transverse position within a small band of frequencies (normalized to a frequency interval of 1 Hz) over a range of frequencies determined by the sampling interval and the overall duration of the recording. In our case, the spindle position was measured in each of 4096 consecutive images acquired at a rate of 31 or 33 frames/ s (depending on the model of the camera) over ≈128 s, corresponding to a range of frequencies from ≈8 mHz to 16 Hz. The time trace of the transverse position of the spindle in Figure 1 is shown at an expanded scale in Figure 2A. The power spectra from the eight embryos had similar amplitudes over the whole frequency range and were averaged (Figure 2B, black line; green circles correspond to the cell in Figure 1). At high frequencies there is an asymptote of about 40 nm$^2$/Hz, the power then climbs with a maximum slope of about 2.5 on the log-log plot as the frequency decreases and then the slope decreases again at low frequency, consistent with a low-frequency asymptote of about 10$^6$ nm$^2$/Hz.

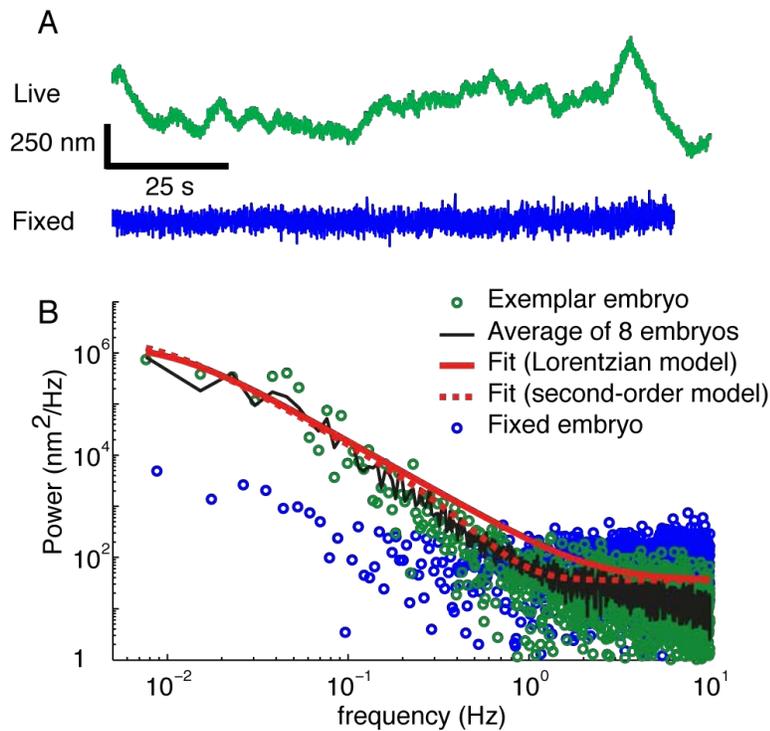

**Figure 2: Power spectrum of the transverse spindle position in control embryos**
**A** Time traces of live and fixed cells. The green, upper curve is the transverse position of the spindle center of the embryo in Figure 1 during the maintenance phase. The blue, lower curve is the spindle transverse position in a methanol-fixed embryo. Note that there is more high-frequency noise in the fixed cell due to the reduction in intensity of the GFP. However, the lower-frequency, biological noise is clearly less in the fixed cell. **B** Experimental and theoretical power spectra. The green circles are the one-sided power spectral density of the *y*-component of the spindle center computed during the maintenance phase. The black line is the average of power spectra from 8 embryos. The blue circles are the power spectrum of the fixed embryo. The solid red line is the least-squares fit to the Lorentzian model with $\sigma^2 = 24.0 \pm 6.0 \times 10^3$ nm$^2$, $\tau = 14.5 \pm 3.8$ s and high-frequency asymptote $\sigma_0^2 = 36$ nm$^2$/Hz. The dashed red line is the least-squares fit to the second-order model with $\sigma^2 = 27.1 \pm 8.4 \times 10^3$ nm$^2$, $\tau = 18.1 \pm 5.7$ s, $\tau_0 = 0.37 \pm 0.02$ s and $\sigma_0^2 = 36$ nm$^2$/Hz.

To prove that these fluctuations are real and not due to measurement noise from the tracking algorithm, we imaged a methanol-fixed embryo and computed the power spectrum of the transverse position of the spindle center (time trace in Figure 2A, spectrum in Figure 2B both shown in blue). At low frequencies, the power was one to two orders of magnitude less than that of the living embryos, showing that the fluctuations are of biological origin. Above 1 Hz, the power measured in the fixed cell was somewhat higher than that in the live cell (also seen in the time trace) due to the reduction in GFP signal following fixation.

**Modeling the dynamics of the fluctuations**

To gain insight into the molecular origin of the fluctuations, we fit two different theoretical curves to the power spectra, a Lorentzian and a second-order model (26). The Lorenzian is the prediction of a model that assumes that the fluctuations are due to a random process with a correlation time, $\tau$; the correlation leads to low pass filtering of the power spectrum with a slope of -2 on a log-log axis at high frequencies. A physical interpretation of this model is that the centering mechanism acts like a spring, which moves the spindle back towards the center, combined with a viscous element, which slows down the movements (25). The correlation time is the drag coefficient of the viscous element ($\gamma$) divided by the spring constant ($\kappa$): $\tau = \gamma / \kappa$. The stiffness and and drag coefficient were recently measured using magnetic tweezers (15). Fluctuations arise from stochastic variation in the number of force generators acting on the spindle. The second-order model has two characteristic times: the longer time constant likely corresponds to a damped spring, as in the Lorentzian model, and the shorter time constant might arise from an active motor-driven process or a relatively fast mechanical process such as microtubule buckling (see Discussion). For both models, we added a frequency-independent noise corresponding to the high-frequency asymptote.

The Lorentzian provided a good fit to the average power spectrum of the transverse position, except in the frequency range between 0.1 and 1 Hz, where the data fell below the theoretical curve (Figure 2A, B, solid black curves). The best-fit Lorenzian parameters were a time constant $\tau = 14.5 \pm 3.8$ s, an estimated total power (over all frequencies) of $24,000 \pm 5,900$ nm$^2$, and a high-frequency asymptote of 36 nm$^2$/Hz (errors in the fits correspond to standard errors, the high-frequency asymptote was set equal to the average of the data over 1-3 Hz). Thus, our data are consistent with a centering machinery that acts as a damped spring, as was inferred by the application of external forces to the spindle (15). The square root of the estimated total power is $155 \pm 19$ nm (mean ± SE). By comparison, the square root of the mean variance of the eight time traces, i.e. the average standard deviation of the time traces, was $131 \pm 12$ nm (mean ± SE, $N = 8$). The similarity between the square root of the estimated power and the average standard deviation indicates that in these embryos the bandwidth of the measurement (from 0.08 to 16 Hz) was great enough to capture most of the variance.

The primary limitation of these measurements is the duration of the maintenance phase (130 seconds), which is less than ten times the time constant. As a result, the low frequency asymptote is not well constrained. Furthermore, if there is a persistent drift, as was observed in some embryos, then there will be no asymptote. In the eight embryos that we analyzed in detail and which showed little drift, the 95% confidence range of the total estimated power is 12,300 - 43,000 nm$^2$, (doing the analysis on the logarithm of the parameter values and using $t = 2.37$ for 7 degrees of freedom); this range is approximately a factor of two on either side of the mean. Thus, the error in the estimated total power is large.

The second-order model provided a good fit to the average power spectrum over the whole frequency range (Figure 2A, B, dashed curve). The parameters were: $\tau_0 = 18.1 \pm 5.7$ s, $\tau_1 = 0.37 \pm 0.02$ s, an estimated total power (over all frequencies) of $27,200 \pm 8,300$ nm$^2$ and a high-frequency asymptote of 36 nm$^2$/Hz as before.

The Fourier analysis confirms that the stability of centering is high. This confirmation is important because it allows to estimate the total fluctuations. We estimate that the standard deviation of the transverse fluctuations divided by the length of the minor axis of the cell is only 0.48% (155 nm/32.5 μm). With 95% confidence, we estimate the SD/axis length is < 0.7%.

For the fluctuations along the AP axis, there was no strong evidence for a low-frequency asymptote. This indicates that the correlation time was longer than 20 s, and the total estimated power was larger than that of the transverse fluctuations, in agreement with the filtered standard deviations (see above).

**Dependence of spindle fluctuations on cortically generated forces**

As noted earlier, towards the end of metaphase the posterior centrosome begins to move towards the posterior (Figure 1D, $P_x$), leading to a posterior displacement of the spindle center (Figure 1E, $S_x$). This posterior displacement sets up an asymmetric cell division, giving rise to the anterior AB and the posterior P1 daughter cells. Posterior spindle displacement is driven by dynein motors attached to cortex that pull on the astral microtubules (32-35). While the motor activity of cortical dyneins is not required for the initial centration of the spindle (e.g. (26)), the cortical dyneins have been proposed to contribute to the maintenance phase (36, 37). We therefore tested whether cortical dyneins contribute to the stability of centering during the maintenance phase.

To test the role of the cortical dynein in spindle stabilization, we partially knocked down, using RNAi, the proteins GPR-1 and GPR-2, which are in a G-protein pathway that controls the motor activity of the cortical dyneins (32). To exclude nonspecific effects of RNAi, we knocked down *lov-1*, a gene that has no role in cell division and found no difference from control. Simultaneous RNAi against both GPR-1 and GPR-2, abolished spindle oscillations (Figure 3A, $S_y$) and delayed asymmetric spindle positioning (Figure 3A, $S_x$), confirming the partial decrease in activity (26, 32, 38-40). Interestingly, the stability of centering increased. The fluctuations in the frequency range 0.1 to 1 Hz decreased by $19 \pm 5$ % along the transverse axis ($p = 2 \times 10^{-4}$) and $46 \pm 6$ % along the AP axis ($p = 7 \times 10^{-9}$) (Table 2). The power spectrum decreased relative to control at all frequencies (Figure 3C, black circles compared to the red line), with an integrated power of $11,500 \pm 3,600$ nm$^2$, about 48 % of the wildtype power. This result shows that the high stability of centering does not require GPR-1/2 dependent pulling forces; indeed, the increase in stability following RNAi indicates that the pulling forces generated by the cortical dyneins have a destabilizing activity.

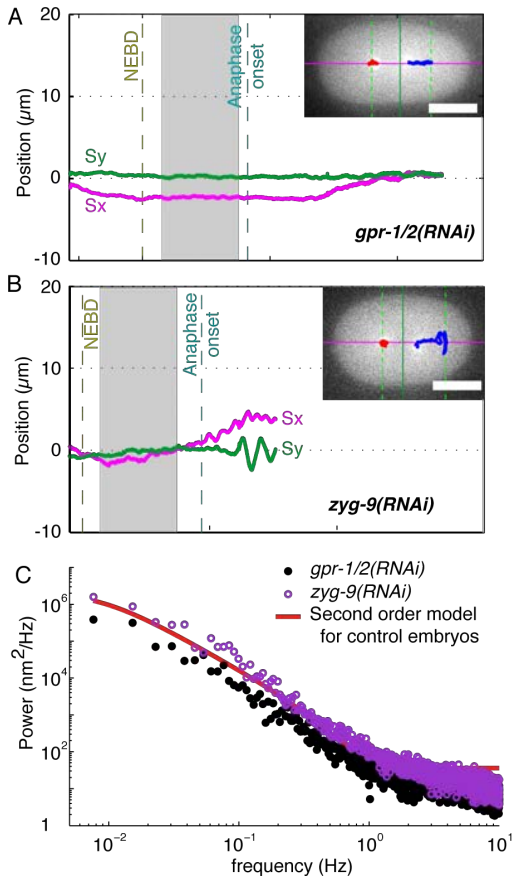

**Figure 3: Spindle positioning in *gpr-1/2(RNAi)* and *zyg-9(RNAi)* embryos**
**A** Time traces of spindle position in a *gpr-1/2(RNAi)* embryo along the AP-axis (magenta) and the transverse (green) axis showing the loss of transverse oscillations. **B** Time traces of spindle position in a *zyg-9*(RNAi) embryo along the AP-axis (magenta) and the transverse (green) axis. **C** Average power spectra of 8 *gpr-1/2(RNAi)* and 8 *zyg-9(RNAi)* embryos (black closed and purple open circles respectively). For comparison, the second-order model fit to the control embryos (From Figure 3) is shown in red. See also Supplementary Figure S1 for positioning in *lin-5(RNAi)* and *nmy-2(RNAi)* embryos.

The GPR-1 and GPR-2 proteins are functionally redundant as the mitotic spindles in the individual mutants undergo both posterior displacement and spindle oscillations (38). In order to study their individual roles in spindle stability, we crossed these mutants into the γTUB::GFP strain. The *gpr-1(ok2126)* mutant had normal posterior displacement and spindle oscillations, whereas the *gpr-2(ok1179)* mutant had normal posterior displacement but reduced spindle oscillations, indicating a weak phenotype. The transverse fluctuations in the frequency range 0.1 to 1 Hz increased by $35 \pm 9$ % ($p= 3\times10^{-4}$) in *gpr-1(ok2126)* and by $11 \pm 10$ % in *gpr-2(ok1179)*, while the AP fluctuations decreased by $6 \pm 9$ % in *gpr-1(ok2126)* and by $17 \pm 8$ % (p=0.05) in *gpr-2(ok1179)* (Table 2). The power spectra were similar to controls (data not shown). Thus, deleting the GPR proteins individually did not have a consistent effect on the fluctuations, increasing the transverse fluctuation but decreasing the axial fluctuations.

RNAi against the NUMA homolog LIN-5, which couples GPR-1/2 to dynein (41), also abolishes spindle oscillations and delays asymmetric spindle positioning (Supplementary Figure S1A). In addition, spindle orientation was delayed. The fluctuations of the transverse position of the spindle center increased by 2 ± 15 % along the transverse axis and decreased by 34 ± 9 % along the AP axis ($p = 0.001$) in the frequency range 0.1 to 1 Hz (Table 2). The power spectrum of the transverse fluctuations decreased slightly (Supplementary Figure S1A). This provides additional evidence that pulling forces are not necessary for the high stability of centering.

GPR-1/2 dependent pulling force generation at the cell periphery depends on the non-muscle myosin NMY-2 (42). We weakened the cortex by partial *nmy-2(RNAi)* and found that centering still occurred (Supplementary Figure S1B). This further argues against pulling forces being required for centering. However, unlike the GPR-1/2 knockdown, there was an increase in the fluctuations in the frequency range 0.1 to 1 Hz: the transverse fluctuations increased by 101 ± 22 % ($p = 5 \times 10^{-5}$) and the axial ones increased by 19 ± 13 % (Table 2). The increase in fluctuations in *nmy-2(RNAi)* shows that the cortex facilitates centering.

**Dependence of spindle fluctuations on microtubule dynamics**

The forces that maintain the spindle at the cell center depend on microtubules (15). To determine whether the stability of centering also depends on microtubules, we slowed down microtubule growth by a mild, though penetrant, knockdown of ZYG-9, the *C. elegans* member of the chTOG/XMAP 215/STU2 family of microtubule polymerases (43). *zyg-9(RNAi)* decreased the growth rate slightly, but significantly, from 0.73 ± 0.03 ($N = 52$ microtubules) to 0.65 ± 0.02 μm/s ($N = 62$ microtubules, $p = 0.05$). There was also a small, but significant, decrease in the number of microtubules arriving at the cortex, 31% ($N = 4$ embryos, $p = 0.009$). Associated with this reduction in microtubule number and growth speed, the stability of spindle centering decreased: while the decrease is difficult to see in individual traces (e.g. Figure 3B), the average amplitude of the fluctuations in the frequency range 0.1 to 1 Hz increased, by 75 ± 28 % in the transverse axis ($p = 0.01$) and 15 ± 15 % ($p > 0.05$) in the AP axis, and the angular fluctuations increased 73 ± 70 % ($p > 0.05$) (Table 2). The power spectrum showed an increase over control at all frequencies (Figure 3C purple circles compared to the red line). Thus, slowing microtubule growth correlates with a decrease in stability of centering.

# Discussion

Our main finding is that the centering of the mitotic spindle in the one-cell *C. elegans* embryo is highly precise and stable. The cell-to-cell variability (i.e. precision) in the position of the spindle center transverse to the anterior-posterior axis, as measured by the standard-deviation, was 470 nm, corresponding to only 1.5% of the length of the short axis of the cell. The transverse fluctuations had an average standard deviation of 155 nm, corresponding to only 0.5% of the length of the short axis of the cell, indicating high stability. The precision and stability were less along the AP axis, but still very high. The high precision is similar to the 1% precision of the hunchback protein profile in the cycle-14 *Drosophila* embryo (44), one of the most precise developmental events studied (45) . The high stability implies that the number of molecules involved in centering must be large. If molecules such as microtubules or motor proteins act independently and are of number *n*, then the relative fluctuation in the number is $1/\sqrt{n}$. Thus, for a precision and stability on the order of 1%, we require on the order of 10,000 centering molecules.

**Comparison of results to centering models**

We now discuss whether our findings are consistent with three centering models that have been discussed in the literature (Figure 4) (46, 47).

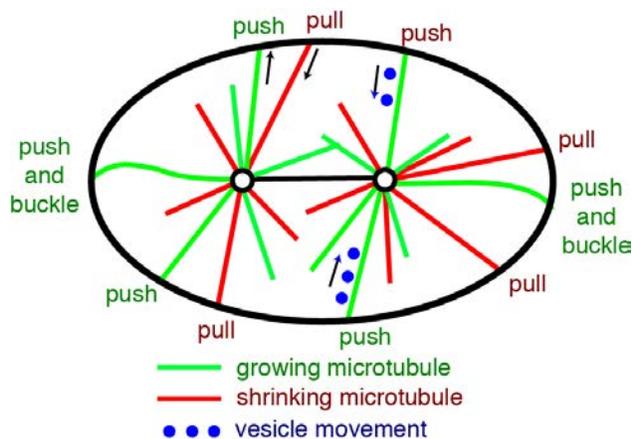

**Figure 4: Models of centering**
Diagram of a one-cell embryo showing microtubules growing from (green) and shrinking to (red) the centrosomes (circles). If a microtubule continues to grow when it contacts the cortex (the inside of the ellipse) then it will push. If the centrosome is closer to one side, the microtubules on that side will spend less time growing and shrinking (because they do not have to go as far) and so will spend a larger fraction of time pushing: this leads to a centering force. If a microtubule shrinks while still in contact with the cortex, then it will pull. Pulling is often destabilizing, though under some circumstances it can lead to centering. If vesicles are carried by motors towards the centrosome, then the drag force on the vesicle will lead to a reactive force on the centrosome and spindle: if the spindle is displaced there will be a net force pulling the centrosome towards the center. Buckling microtubules are shown at the ends: the left one cannot slide on the cortex; the right one can slide.

(i) <u>Cortical pulling.</u> Astral microtubules grow out to the cell periphery where they interact with cortical-anchored motors that generate pulling forces (48, 49). This cortical force generator activity, of which cytoplasmic dynein is an essential component, can be greatly reduced by RNAi against *gpr-1/2* and other proteins that define a regulatory pathway (26, 32, 50). When these proteins are knocked down, the spindle still centers, showing that this pathway is not essential for the initial establishment of centering. Our results show that the cortical force generators are not necessary for the high stability of centering during the maintenance phase and indeed reducing force-generator activity leads to an increase in stability. These findings are consistent with a recent study showing that knocking down GPR-1/2 increases the force associated with centering (15). Together, these results and ours suggest that the cortical force generators have an anti-centering activity. Such anti-centering activity is expected because the closer the centrosome to the cortex, the larger the net pulling force (25). Thus, cortical pulling does not stabilize centering.

(ii) Cytoplasmic pulling. Membrane-bound organelles move along the astral microtubules towards the centrosomes. The viscous forces acting on the organelles lead to a reactive force that moves the spindle towards the moving organelles (8, 9, 51). If the spindle is displaced from the cell center, the astral microtubules will be longer on one side than the other, and the reactive force will tend to move the spindle back to the center (9, 10, 30, 52). Thus, cytoplasmic pulling forces are centering.

The cytoplasmic pulling model accords with most our observations. The model relies on hydrodynamic forces generated by vesicle movement. Given that there are many thousands of microtubules per centrosome (see below) and that several vesicles can potentially move on each microtubule (9, 53), the number of moving vesicles may be large enough to attain the high observed stability. In addition, given that movement of the centrosome entails a re-equilibration of the distribution of microtubules, and that this is likely to take on the order of the times to grow to and shrink from the cortex (on the order of 20 s each way, see next paragraph), the correlation time of the fluctuations is also consistent with cytoplasmic pulling. The cytoplasmic pulling model is not readily consistent with the *nmy-2* RNAi because centering by this mechanism is not expected to be influenced by activity at the cortex. However, an indirect effect of the cortex on vesicle transport could affect centering. Thus, our data do not rule out the cortical pulling model.

(iii) Cortical pushing. There are two lines of evidence that microtubules continue to grow after contact with the cortex (and therefore generate pushing forces): (i) in vivo, EB1, a marker for microtubule growth, continues to bind to microtubule ends after they contact the cortex (23). (ii) in vitro, microtubules fixed at one end buckle when the other end makes contact with a solid surface (24), indicating that growth continues and that compressive (pushing) forces are generated. Astral microtubules will spend a larger fraction of their time pushing on the side closer to the cortex, because the microtubules spend less time growing to and shrinking from the cortex. This leads to a force imbalance that tends to return the spindle to the center of the cell (25). Thus cortical pushing forces lead to centering.

Microtubule pushing is consistent with several properties of centering. First, it accords with the small forces associated with spindle centering, ~16 pN per 1 μm displacement from the cell center (15). We note that the centering stiffness, 16 pN/μm, implies that the Brownian motion is only 16 nm (standard deviation over the full bandwidth of the fluctuations), which is about 10-fold smaller than the estimated fluctuation in wild-type cells of 155 nm. Second, pushing accords with the greater stability along the shorter transverse axis than along the longer AP axis (Table 1): the fluctuations are expected to be smaller in smaller cells because the growth and shrinkage times of microtubules from centrosome to cortex are shorter (25). Third, the decrease in stability following *nmy-2* RNAi is consistent with a cortical mechanism such as pushing, though, as pointed out before, the effect of *nmy-2* knockdown may be indirect. Fourth, pushing is known to center and orient the mitotic spindle in other cells, such as fission yeast (54, 55). It has, however, been argued that in large metazoan cells, such as those in the *C. elegans* zygote, the microtubules are likely to buckle (two examples of buckling microtubules are depicted in Figure 4) and that the associated reduction in pushing force will make pushing an inefficient centering mechanism (46); we address this below. Thus, several observations support the pushing model.

One observation that is difficult to reconcile with the pushing model is that the high stability of centering appears to be at odds with the comparatively small number of pushing microtubules. Even though there are at least 2000 microtubules per spindle pole (56), the fraction that are in contact with the cortex at any one time is low because the interaction with the cortex is transient. If there are $M$ microtubules on each side of the spindle and they are pushing (without buckling) for a fraction $p_0$ of the time, then the cortical pushing model predicts that the SD of the fluctuations divided by the cell diameter is $\approx 1/(2\sqrt{Mp_0})$ (assuming $p_0$ is small as is the case because the pushing times are much shorter than the growing and shrinking times, see Materials and Methods and (25)). We estimate from data obtained by imaging microtubule ends at the cortex *en face* ((15), Supplementary Figure 10B and using the surface area in Supplementary Figure 8E) that during metaphase, the total number of microtubule ends marked with EB1 interacting over the entire cortex at any one time, $Mp_0$, is 236 (only scoring microtubules ends with interaction times >0.4 s). The interaction time distribution was exponential with a time constant of 0.7 s (the average of the scored times was 0.99 s). Correcting for missed interactions (i.e. those shorter than 0.4 s), we estimate the number of cortex-interacting microtubules to be 418. This number is similar to that inferred from (23) and our own independent measurements obtained by SR. Using this value, the cortical pushing model predicts that the standard deviation of the fluctuations divided by centrosome-cortex distance is 2.4%. This is not consistent with the measured value of 0.5% and the 95% confidence bound of 0.7% (and about 25% less in GPR1/2 RNAi cells). Thus, the number of pushing microtubules appears to be too small to account for the high stability.

However, if the microtubules buckle, as expected based on *in vitro* experiments (57, 58), then the stability is expected to increase (not decrease as had been assumed (45)). This is because of the length-dependence of buckling: the microtubules on the shorter side reach a larger force before they buckle, leading to stronger centering. In this case (25), the relative stability is increased approximately three-fold to $\approx 1/(6\sqrt{Mp_0}) \approx 0.8\%$. The key point here is that the relative stability is independent of the magnitude of the pushing force; this is because positional fluctuations due to thermal forces are very small, given the stiffness of the centering apparatus (15). For this reason, the diminution of pushing forces by buckling does not degrade stability, as had been thought. Thus, the measured stability is close to, though a little smaller, than the stability predicted by the cortical pushing model.

Microtubule buckling also accounts for the measured time constant of centering. The time constant arises from the re-equilibration of the microtubule array and depends on the time that it takes for the microtubule to grow out from the centrosome and shrink back from the cortex. In the case of buckling, the time constant is expected to be $\tau \approx \frac{1}{3}(\tau^+ + \tau^-) \approx 16$ s (25), where the growth and shrinkage times are respectively $\tau^+$ = 20.5 s and $\tau^-$ = 17.9 s, assuming a distance of 15 μm from the nucleation site in the centrosome to the cortex and growth rate of 0.73 μm/s (43) and a shrinkage rate of 0.84 μm/s (23). The time constant is on the order of the time that it takes for a microtubule to grow from the centrosome to the cortex and shrink back again. The predicted time constant is consistent with the measured time constant of 14.5 ± 3.8 s for the Lorentzian model. In the absence of buckling, the time constant is expected to be more than ten times longer, which is inconsistent with the data (25). Thus, the pushing model accords with the data provided that the microtubules buckle. As an aside, the data in the last few paragraphs can be used to estimate the total number of astral microtubules in a cell. The fraction of the time microtubules were observed at the cortex was 0.99/(20.5 + 17.9 + 0.99) = 0.025. Assuming that there are no catastrophes or rescues in the cytoplasm, we can estimate that the total number of astral microtubules is ≈9,000 (236/0.025). This is about twice as high as a lower estimate of total microtubule number of 4000 based on light and electron microscopy (56).

Thus, buckling can account for the measured stability. It is important to point out, however, that buckling has only been observed during anaphase when the spindle oscillates (23, 43). In this case, the force that drives buckling may originate from the cortical dyneins that drive the oscillations (26). Buckling has not been observed during the quiescent maintenance phase, and thus stabilization by buckling remains hypothetical.

**The kinetics of centering**

Our results on the kinetics of spindle fluctuations are in general agreement with the magnetic tweezer experiments of (15). They found that response to force, the spindle displayed viscoelastic behavior with a spring constant of $\kappa$ = 16 pN/μm and a drag coefficient of $\gamma$ = 130 pN·s/μm. The time constant, $\gamma/\kappa \approx 8$ s, is within a factor of two of our correlation time of 15 ± 4 s. The uncertainty in the correlation time is large because the total maintenance phase is less than ten times longer than the correlation time; for this reason, the difference between the correlation time and the time constant measure in the force experiments is not significant.

A final discussion point is our finding that the Lorentzian model did not provide a good fit to the power spectra. A discrepancy was observed in the mid-frequency range between 0.1 and 1 Hz where the power decreased more rapidly with increasing frequency than predicted. The more rapid decrease implies that the autocorrelation function is not a single exponential. The good fit using a second-order model indicates that the autocorrelation is well fit with two exponentials. It is reasonable that the longer-time-constant exponential corresponds to the re-equilibration of the microtubule array due to growth and shrinkage, as is the case for the Lorentzian. We are unsure what the second, shorter time constant corresponds to. One possibility is that it corresponds to the active process that drives oscillations: the active process is expected to have two time constants and to be present even before the oscillation threshold is reached (26). Other possible explanations for the second time constant include microtubule buckling, which happens on a fast time-scale (59), or delays associated with cortical catastrophe. Mixed pushing-pulling models have also been proposed (37, 60), and these may lead to second-order kinetics. Further work will be required to test these possibilities.

**Conclusion**

Using high-resolution tracking and Fourier analysis of spindle trajectories, we found that the accuracy, precision and stability of spindle positioning during metaphase of mitosis of the one-cell embryo of C. *elegans* is very high. The maintenance of spindle position during metaphase could not be accounted for by microtubule pulling by cortical motor because depletion of the cortical pulling force generators resulted in improved, not diminished, centering. Of other possible centering mechanisms, microtubule pushing against the cortex, taking advantage of buckling to create additional stabilizing feedback, accounts for the high stability of centering, though other models, such as cytoplasmic pulling cannot be ruled out by our data.

## Author contributions

JP, SR, AAH and JH designed research; JP, SR, ZA, BM, SP and SG-C performed research; JP and JH analyzed data; JP and JH wrote the paper.

## Acknowledgements

The authors thank Dr C. Kozlowski & Dr F. Nedelec for providing their landing assay algorithm, Drs C. Brangwynne, H. Bouvrais, L. Chesneau & Y. Le Cunff for comments on the manuscript and Drs H. Fantana, V. Bormuth, G. Greenan & E. Schäffer for stimulating discussions. J.P. was supported by the Human Frontier Science Program Organization. The PécréauxLab was funded by the CNRS/INSERM and "Ligue Nationale Contre le Cancer, LNCC" (ATIP/Avenir) and by the "Région Bretagne" (grant AniDyn). B.M. was supported by a LNCC PhD fellowship. JH was supported by the Max Planck Society and Yale University. We thank an anonymous reviewer of an earlier draft of this manuscript for detailed criticisms which led to a substantial improvement in the analysis of the data.

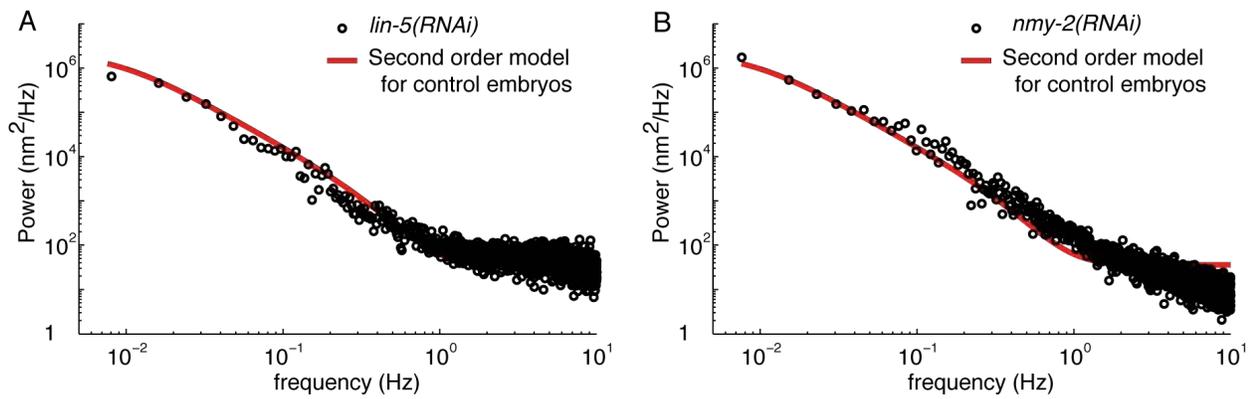

**Figure S1, related to Figure 3: Spindle positioning in *lin-5(RNAi)* and *nmy-2(RNAi)* embryos**

**A** Average power spectra of the transverse spindle position for six *lin-5(RNAi)* embryos (black open circles). For comparison, the second-order model fit to the control embryos (From Figure 2) is shown in red. **B** Average power spectra of the transverse spindle position for seven *nmy-2RNAi)* embryos (black open circles). For comparison, the second-order model fit to the control embryos (From Figure 2) is shown in red.